# Large-amplitude oscillation dynamics and domain suppression in a superlattice Bloch oscillator

Herbert Kroemer*

ECE Department, University of California, Santa Barbara, CA 93106

**Abstract**

We analyze the current dynamics of a superlattice Bloch oscillator under conditions where, in addition to the dc bias field causing a negative differential ac conductivity, a similarly strong ac field is present, at a frequency somewhat below the Bloch frequency associated with the dc field. The differential conductivity at the ac drive frequency is then negative, but, because of the strong non-linearities of the system, the dc differential conductivity is modified by the ac field. For a sufficiently strong ac field, there will be a dc bias range inside which the dc differential conductivity has turned positive, while the ac conductivity remains negative. The phenomenon should lend itself to the suppression of the domain-like space charge instabilities that are normally associated with a negative dc differential conductivity.

## 1. Introduction: The problem

It has been known since the groundbreaking 1970 paper by Esaki and Tsu [1] that a semiconductor superlattice (SL) biased by a strong electric field may, under certain favorable conditions, exhibit a negative *differential* conductivity (NDC) up to frequencies on the order of the Bloch oscillation frequency. This makes the SL a potential candidate for the active medium in a Bloch oscillator (BO) operating all the way up into the terahertz regime.

An important next step was taken in a 1971 paper by Ktitorov, Simin, and Sindalovskii (KSS) [2], who worked out the *small-signal* complex conductivity $\sigma(\omega)$ of a semiconductor SL under conditions of an electron relaxation time sufficiently long and a dc bias field $E_0$ sufficiently high that Bloch oscillations play a key role in the electron dynamics. As is shown in Fig. 1, the real part of the complex conductivity—already negative at zero frequency—will become more negative with increasing frequency, until it reaches a resonance minimum at some frequency not far below the Bloch frequency. (For a detailed discussion of the physical origin of this resonance, see [3].) Presumably, this resonance regime would be the proper frequency regime for a BO.

More recently, there have been several theoretical studies of some aspects of the *large-signal* current dynamics, when the ac current is no longer linear in the ac component of the overall bias field. Most relevant to the goal of achieving a BO as an actual THz power source is the work by Ignatov et al. [4, 5].

* kroemer@ece.ucsb.edu

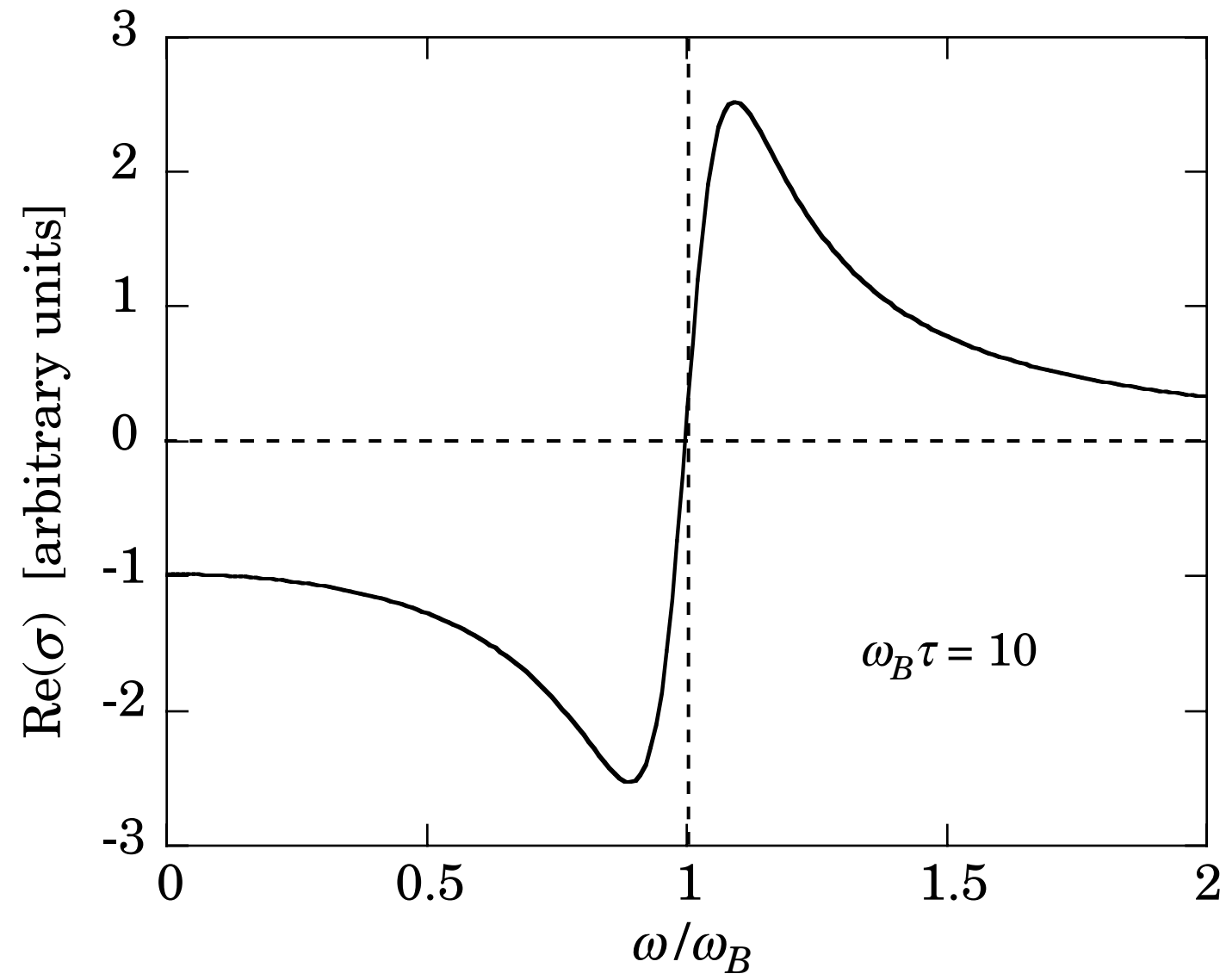

**Fig. 1.** Real part of the KSS differential conductivity vs. frequency, assuming $\omega_B\tau = 10$, where $\omega_B$ is the (angular) Bloch frequency—defined in (4)—and $\tau$ the electron relaxation time, assuming equal momentum and energy relaxation times. From [3].

However, experimental progress towards this goal has been elusive. Only in recent years have there been reports of *transient* true Bloch oscillations following optical pulse excitation (see, for example, [6, 7]; for up-to-date references, see [8]). But I am not aware of any reports on true continuous-wave (cw) Bloch oscillations. Although there *are* several reports on cw oscillations in superlattices (see, for example, [9, 10]), all these oscillations are at much lower frequencies; they are basically transit-time oscillations similar to Gunn-effect oscillations, caused by domain instabilities in a medium in which the negative differential conductivity extends down to zero frequency.

That last point is central. For a true BO to operate in a cw mode, the existence of an NDC just below the Bloch frequency is a first necessary condition; it is also necessary that any negative conductivity at *low* frequencies be suppressed: If, in a *bulk* NDC medium, the NDC persists down to zero frequency, a uniform internal electric field (implying zero space charge) will not be stable, but regions of charge accumulation and/or depletion will develop, leading in turn to regions with different internal fields, similar to the well-known domains in the Gunn effect. These domains in turn tend to suppress the negative *overall* conductance of the device at the intended oscillation frequency just below the Bloch frequency. (For an elementary review of this Gunn effect background, see [11]; detailed reviews are found in [12] and [13]. Much of the space charge phenomenology of the Gunn effect carries over to the BO, despite the very different physics causing the NDC itself).

Oscillations may still take place even in the presence of domains, but they are invariably at or near the transit time frequency $f_t = v/L$, where $v$ is the propagation speed of the domains, and $L$ the length of the sample. Whatever interest these transit-time oscillations may have as physics exercises, for device applications their frequencies are too low to be competitive with mainstream transistor oscillators, and from a BO perspective they are a nuisance, which may in fact have been responsible for the failure—so far—to achieve true cw Bloch oscillations. Hence the suppression of the NDC at low frequency is a key task if one wishes to utilize the NDC at high frequencies in a true Bloch oscillator.

One promising approach towards domain suppression has been proposed by Allen [14], who has pointed out that the KSS resonance enhancement should make it possible to suppress any domain instabilities induced by any NDC at low frequencies, by simply shunting the SL layers, layer-by-layer, with a positive conductance that is sufficiently high to make the differential dc conductivity positive, without obliterating the NDC just below the Bloch frequency. He also pointed out that such operation calls for a dc bias field far above the critical field $E_c$.

In the present paper, we pursue an alternative *dynamic* scheme for domain suppression, similar to the so-called LSA mode (= **L**imited **S**pace charge **A**ccumulation mode) of the Gunn effect [11-13]. The central idea behind the LSA mode is that space charge instabilities may be suppressed if the ac part of the drive field is kept so strong that the overall field dips, during each cycle, to very low values at which the *static* velocity-field-characteristic has a steep positive slope (see Fig. 2 below). Under steady-state operation at such low fields, the medium would be an "ordinary" conductor with a positive conductivity, and any space charges would decay rather than build up. Given a suitable combination of the dc and ac operating conditions, domains are then unable to build up.

As I shall demonstrate in the present paper, a similar dynamic stabilization should also be achievable in a BO. In fact, the same phenomena that cause the KSS resonant NDC enhancement visible in Fig. 1 also facilitate the dynamic stabilization. However, the details are significantly different from the LSA case, reflecting the underlying differences in the physical mechanisms responsible for the NDC.

The dynamic stabilization problem is a special case of the more general question of the electron dynamics when the high-frequency component of the drive field builds up to values that are no longer small compared to the dc component. This is a central problem in the circuit theory of *all* negative-conductance oscillators: When a negative conductance is inserted into a high-$Q$ resonance circuit, any oscillations at the resonance frequency $\omega$ will not decay, but will build up until their amplitude becomes limited by the non-linearities of the negative conductance. With increasing amplitude, the negative-conductance gain will decrease, and the oscillations will saturate at that level at which the negative-conductance gain equals the circuit losses, including the "losses" due to power being extracted from the circuit. The exact amplitude at which this will take place is a problem in non-linear circuit theory, outside the scope of our investigation; we will simply view the ac amplitude as given.

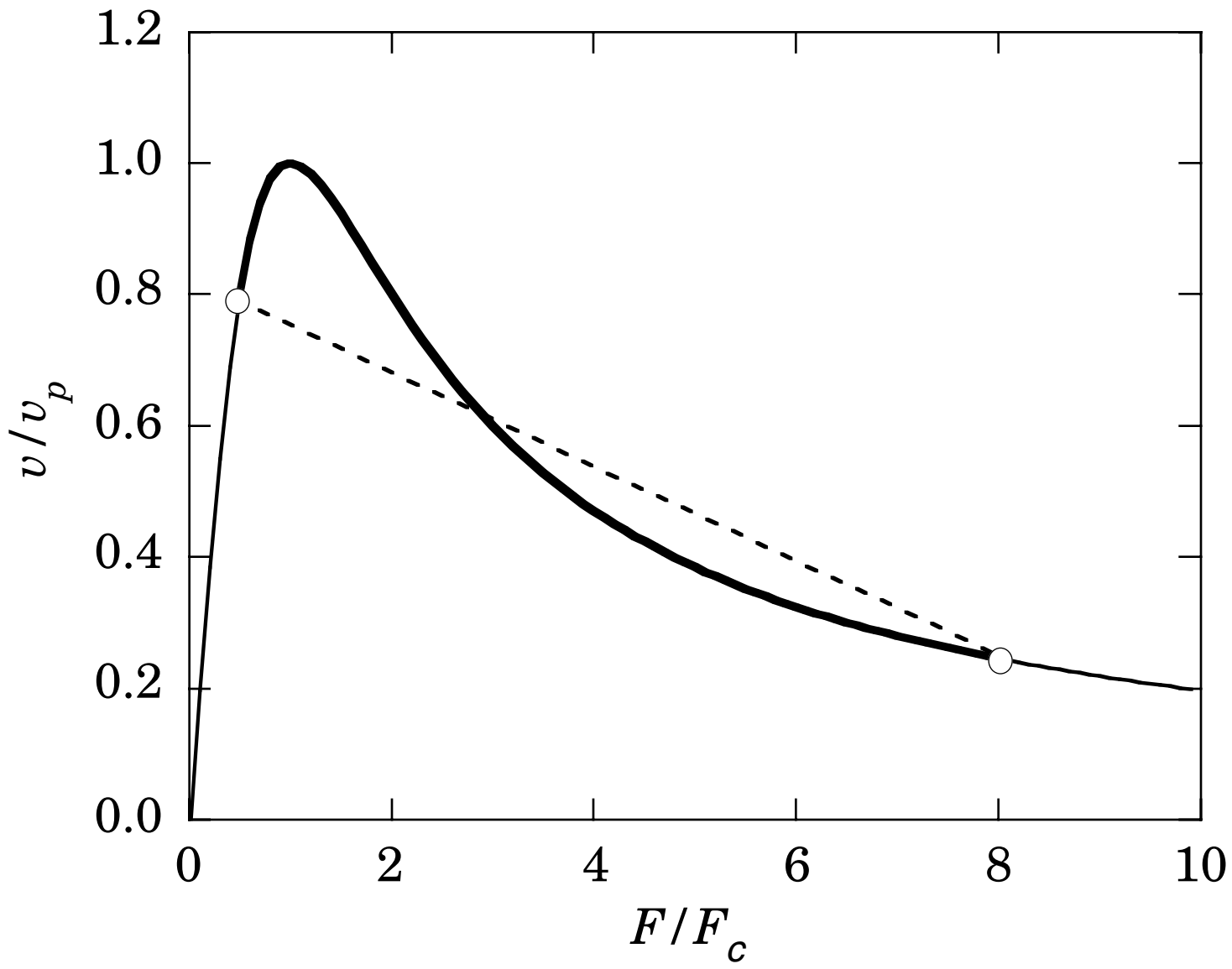

**Fig. 2.** Static drift-velocity-vs.-field characteristic, in normalized units ($v_p$ is the peak drift velocity, and $F_c$ is the critical field at which the peak occurs). The heavy portion of the curve illustrates LSA-type operation in the Gunn effect; it has a much steeper slope at the low-field end than at the high-field end. Current and field oscillate along this portion; a load line (broken line) with a negative slope ensures power delivery to the external circuit. For the form of $v(F)$, see (25) below.

## 2. Distribution function

### *2.1. Preliminaries*

We consider a quasi-one-dimensional SL potential with period $a$, and with a simple electron dispersion relation of the form

$$\mathscr{E}(k) = \tfrac{1}{2}\mathscr{E}_0 \cdot (1 - \cos ka), \tag{1}$$

where $\mathscr{E}_0$ is the miniband width. Associated with this dispersion relation is the electron (group) velocity in real space,

$$v(k) = \frac{1}{\hbar} \cdot \frac{d\mathscr{E}}{dk} = v_{\max} \cdot \sin ka, \ \text{where} \quad v_{\max} = \frac{1}{2\hbar} \cdot \mathscr{E}_0 a. \tag{2a,b}$$

Typical parameters for a SL might be $\mathscr{E}_0 = 30$ meV (just below the optical-phonon energy), $a = 20$ nm, leading to $v_{\max} = 4.56\times10^{7}$cm/s.

In the presence of a spatially uniform applied force $F = -eE$, each electron moves through (extended) $k$-space according to "Newton's law in $k$-space,"

$$\hbar \frac{dk}{dt} = F, \tag{3}$$

where $k$ is a time-dependent *effective wave number* of the electron (see Appendix). In the reduced-zone representation, the electron undergoes an umklapp process whenever it reaches the zone boundary at $k = \pm\pi/a$, re-entering the reduced zone from the opposite boundary.

If only a dc force $F_0 = -eE_0$ acts on the electrons, the (extended) wave number $k$ of an electron increases linearly in time, and the velocity oscillates purely sinusoidally with the (angular) Bloch frequency

$$\omega_B = \frac{1}{\hbar} F_0 a. \tag{4}$$

The corresponding traversal time through the Brillouin zone (BZ) is $t_B = 2\pi/\omega_B$.

Throughout the present paper, we will assume that the overall bias force contains both a dc force $F_0$ and a single high-frequency force with amplitude $F_\omega$,

$$F(t) = F_0 + F_\omega \cdot \cos \omega t. \tag{5}$$

Here the dc term represents an externally applied bias, and the ac term is due to the action of a resonance circuit on the SL.

### *2.2. Rate Equations*

In the limit of an electron scattering frequency that is negligibly small compared to the Bloch frequency, the electrons would eventually become randomly distributed throughout the BZ. We treat $k$ as a continuous variable, and normalize the distribution function such that $f \cdot dk$ is the number of electrons in a $k$-space interval $dk$ wide, per unit real-space volume. The zero-order distribution function in the limit of vanishing scattering is then simply a constant,

$$f_0 = Na/2\pi, \tag{6}$$

where $N$ is the total number of electrons per unit volume (in a true 1-D system, $N$ would be the number per unit *length*). In the present work, we use this uniform distribution as the zero-order perturbation limit, rather than using the thermal-equilibrium limit. Put differently, we view the bias field as part of the unperturbed problem, and the scattering as the perturbation.

In the presence of scattering, the distribution function can always be expanded into a Fourier series in $k$-space. If we restrict ourselves to the lowest-order Fourier coefficients, we may express the distribution function in the simple form

$$f(k,t) = f_0 \cdot [1 + c(t)\cdot \cos ka + s(t)\cdot \sin ka], \tag{7}$$

where $c(t)$ and $s(t)$ are the time-dependent expansion coefficients for the fundamental cosine and sine terms.

Next to our restriction to quasi-one-dimensional dynamics, the neglect of higher $k$-space harmonics is the main approximation of the present paper. To justify it, note that we are principally interested in relatively narrow bands, and in sufficiently high fields and low scattering rates that the scattering frequency is smaller than the Bloch frequency. Under these conditions, the higher $k$-space harmonics should be small, and it is probably an excellent approximation to neglect them altogether.

In (7) we used stationary functions as the basis functions for the Fourier expansion. Such a treatment is well-adapted to the case that the bias field is a pure dc field, in which case the distribution will itself be stationary, with time-independent expansion coefficients $c$ and $s$. But this changes in the presence of an ac component in the bias field. It was shown in [3] that the ac component, combined with inelastic scattering events, induces traveling waves in $k$-space into the distribution function. This suggests using traveling waves as basis functions from the outset.

A useful alternative formulation would be of the form

$$f(k,t) = f_0 \cdot \{1 + C(t)\cdot \cos[ka - \varphi(t)] + S(t)\cdot \sin[ka - \varphi(t)]\}, \tag{8}$$

where the phase $\varphi(t)$ is related to the driving force $F(t)$ via

$$\hbar \frac{d\varphi}{dt} = F(t)\cdot a. \tag{9}$$

The idea behind this formulation is that, in the absence of scattering, every part of the distribution travels through the BZ exactly according to (3). Hence, in that limit, the expansion coefficients $C$ and $S$ become time-independent, and can be chosen arbitrarily, except for the requirement that the distribution function must remain positive.

The two sets of expansion coefficients in (7) and (8) can be expressed in terms of each other,

$$c = C \cdot \cos\varphi - S \cdot \sin\varphi, \qquad s = C \cdot \sin\varphi + S \cdot \cos\varphi, \tag{10a,b}$$

$$C = c \cdot \cos\varphi + s \cdot \sin\varphi, \qquad S = -c \cdot \sin\varphi + s \cdot \cos\varphi. \tag{11a,b}$$

In the zero-scattering limit, the time-independence of $C$ and $S$ leads to the rate equation pair for $c(t)$ and $s(t)$,

$$\frac{d}{dt} c(t) = -s(t) \cdot \frac{d\varphi}{dt}, \qquad \frac{d}{dt} s(t) = +c(t) \cdot \frac{d\varphi}{dt}. \tag{12a,b}$$

To account for scattering, we must add relaxation terms to both equations, yielding the complete rate equations

$$\frac{d}{dt} c(t) = -s(t) \cdot \frac{d\varphi}{dt} - R_c, \quad \frac{d}{dt} s(t) = +c(t) \cdot \frac{d\varphi}{dt} - R_s. \tag{13a,b}$$

We use here the very simple forms for the relaxation rates,

$$R_c = \frac{1}{\tau} \cdot \left[ c(t) - 1 \right], \quad R_s = \frac{1}{\tau} \cdot s(t), \tag{14a,b}$$

with a common time constant for both rates. The ($c$–1)-dependence assumed in (14a) implies that under full relaxation the electron concentration at the top of the miniband just drops to zero, and doubles at the bottom of the band. This is the maximum relaxation consistent with our neglect of higher harmonics in (7). It would be possible to use a limit $c_0 < 1$ in (14a), but the final results would differ only by an uninteresting scale factor.

The use of identical time constants for both rates does not follow the common practice of assuming different relaxation times for energy relaxation and momentum relaxation. But in the presence of inelastic scattering, the relaxation time approximation with an energy-independent time constant is not a very good approximation in the first place. It is then of doubtful relevance to introduce separate energy and momentum relaxation times, especially in 1-D systems, where elastic and inelastic scattering are not distinct processes: every scattering event involves both a momentum and an energy change, with a one-to-one relation between the two. The distinct energy and momentum relaxation times introduced in KSS and other treatments are really not independent quantities, they simply quantify two aspects of a common process. This changes in 3-D (for a discussion, see [3]), but there is little point in introducing this distinction in a 1-D treatment such as ours, where other approximations are far more significant.

If we insert (14) into (13) and transform the result back to the traveling-wave formalism, we obtain two very simple decoupled equations for $C(t)$ and $S(t)$, which may be written

$$\tau \frac{dC}{dt} + C = +\cos\varphi, \qquad \tau \frac{dS}{dt} + S = -\sin\varphi. \tag{15a,b}$$

*2.3. Distribution function under combined dc/ac drive*

Given an overall force of the form (5), the phase $\varphi(t)$ in the rate equations (15a,b) is of the form

$$\varphi(t) = \omega_B t + z \cdot \sin\omega t, \text{ with } \quad z = F_\omega a/\hbar\omega, \tag{16a,b}$$

where we have taken a zero integration constant.

The terms on the right-hand sides of (15) may then be written as standard Fourier-Bessel series,

$$\cos[\varphi(t)] = \sum_{n=-\infty}^{+\infty} (-1)^n \, \mathrm{J}_n(z) \cdot \cos(\omega_n t), \tag{17a}$$

$$\sin[\varphi(t)] = \sum_{n=-\infty}^{+\infty} (-1)^n \, \mathrm{J}_n(z) \cdot \sin(\omega_n t), \tag{17b}$$

where $\mathrm{J}_n$ is the $n$th-order Bessel function, with $\mathrm{J}_{-n} = (-1)^n \mathrm{J}_n$, and

$$\omega_n \equiv \omega_B - n\omega. \tag{18}$$

If (17) is inserted into (15), $C(t)$ and $S(t)$ must be of the form of Fourier-Bessel series of their own. One confirms easily that

$$C(t) = \sum_{n=-\infty}^{+\infty} \frac{(-1)^n \, \mathrm{J}_n(z)}{1 + (\omega_n \tau)^2} \cdot \left[\omega_n \tau \cdot \sin(\omega_n t) + \cos(\omega_n t)\right], \tag{19a}$$

$$S(t) = \sum_{n=-\infty}^{+\infty} \frac{(-1)^n \, \mathrm{J}_n(z)}{1 + (\omega_n \tau)^2} \cdot \left[\omega_n \tau \cdot \cos(\omega_n t) - \sin(\omega_n t)\right]. \tag{19b}$$

*2.4. Current density*

Given a distribution function $f(k)$ of the form (7), and drawing on the velocity-vs.-$k$ relation (2), we obtain the *particle* current density

$$j = \int_{-\pi/a}^{+\pi/a} v(k) f(k) dk = v_{\max} \int_{-\pi/a}^{+\pi/a} \sin ka \cdot f(k) dk$$
$$= v_{\max} \cdot f_0 \cdot s(t) \int_{-\pi/a}^{+\pi/a} \sin^2 ka \cdot dk = \tfrac{1}{2} N \cdot v_{\max} \cdot s(t). \tag{20}$$

The electrical current density is obtained by multiplying with the electron charge $-e$; however, throughout this paper, all currents will be expressed as particle currents. Similarly, we will also often refer to the forces $F$ as fields.

Note that (20) contains no contribution from the cosine term in (7). In fact, if we had included higher $k$-space harmonics, these would also have not made any contributions; their only effect would have been an indirect one, through any change in $s(t)$ introduced by the competition from the higher harmonic terms.

Inserting (19) into (10b), and once more utilizing (17), yields

$$j(t) = \tfrac{1}{2} N \cdot v_{\max} \cdot \sum_{m,n=-\infty}^{+\infty} \frac{(-1)^{m-n} \, \mathrm{J}_m \cdot \mathrm{J}_n}{1+(\omega_n \tau)^2} \cdot \left\{ \omega_n \tau \cdot \cos\left[(n-m)\omega t\right] + \sin\left[(n-m)\omega t\right] \right\}. \tag{21}$$

We evidently have both a dc current (from to $m = n$ terms), and currents at the drive frequency $\omega$ and its harmonics.

## 3. DC conductivity

The dc part of (21) may be written

$$j_0 = \tfrac{1}{2} N \cdot v_{\max} \sum_{n=-\infty}^{+\infty} \left[\mathrm{J}_n(z)\right]^2 \cdot \frac{\omega_B \tau - n\omega\tau}{1+(\omega_B \tau - n\omega\tau)^2}. \tag{22}$$

Consider first the limit of pure dc drive, $F_\omega = 0$. Then, with $z = 0$, all Bessel functions other than $\mathrm{J}_0$ vanish, and (22) reduces to the contribution from the $n = 0$ term:

$$j_0 = \tfrac{1}{2} N \cdot v_{\max} \cdot \frac{\omega_B \tau}{1+(\omega_B \tau)^2}. \tag{23}$$

We re-express this result in terms of the static electron drift velocity $v$, defined via

$$j_0 = N \cdot v(F_0). \tag{24}$$

Evidently,

$$v(F_0) = \tfrac{1}{2} v_{\max} \cdot \frac{\omega_B \tau}{1+(\omega_B \tau)^2} = v_p \cdot \frac{2F_0/F_c}{1+(F_0/F_c)^2}. \tag{25}$$

In the last form, we have drawn on the relation (4) between Bloch frequency and the dc bias force, and have introduced two phenomenological parameters characterizing the velocity-field characteristic,

$$v_p = \tfrac{1}{4} v_{\max} = \mathscr{E}_0 a/8\hbar \quad \text{and} \quad F_c = \hbar/a\tau. \tag{26a,b}$$

Here, $v_p$ is the peak drift velocity, and $F_c$ is the *critical field* at which $v$ goes through this maximum. The $v(F)$ characteristic in Fig. 2 is of this mathematical form.

For a BO, we need $F_0 > F_c$; evidently the dc differential conductivity in the absence of any ac field is negative. However, this changes when an ac field is present. If we re-write (22) in terms of the velocity-field characteristic $v(F)$, we obtain

$$j_0 = N \cdot \sum_{n=-\infty}^{+\infty} \left[ \mathrm{J}_n(z) \right]^2 \cdot v(F_0 - nF^*), \tag{27}$$

where

$$F^* = \hbar\omega/a \tag{28}$$

is that force for which the potential drop across one SL period is equal to the photon energy at the drive frequency $\omega$. We therefore refer to the shifted functions $v(F_0 - nF^*)$ with $n \neq 0$ as *photon replicas* of the static $v(F)$ characteristic, displaced along the field axis by multiples of $F^*$. Note that, for $F_0 = F^*$, the Bloch frequency $\omega_B$ would coincide with the drive frequency $\omega$.

The Bessel function argument $z$, defined in (16a), may be re-expressed in terms of $F^*$ as

$$z = F_\omega/F^*. \tag{29}$$

The full sum in (27) for the dc current density evidently represents a superposition of the velocity-field characteristic and its various photon replicas, each term having an amplitude given by $\mathrm{J}_n^2$. Fig. 3 shows the result of this summation, plotted as a function of the dc field $F_0$, in units of $F^*$, assuming $F^*/F_c = 10$ and $F_\omega/F^* = z = 1$. We shall use this set of *default parameters* frequently throughout the remainder of this paper.

We evidently have a very steep positive slope in the vicinity of $F^*$, representing a positive dc conductivity in this regime. The peak in the curve occurs slightly below $F^* + F_c$. Inasmuch as a negative ac conductivity at the drive frequency $\omega$ requires $F_0 > F^*$, the bias range of interest for BO operation is

$$F^* < F_0 < F^* + F_c, \tag{30}$$

We shall see later that the ac conductivity in this range is in fact negative.

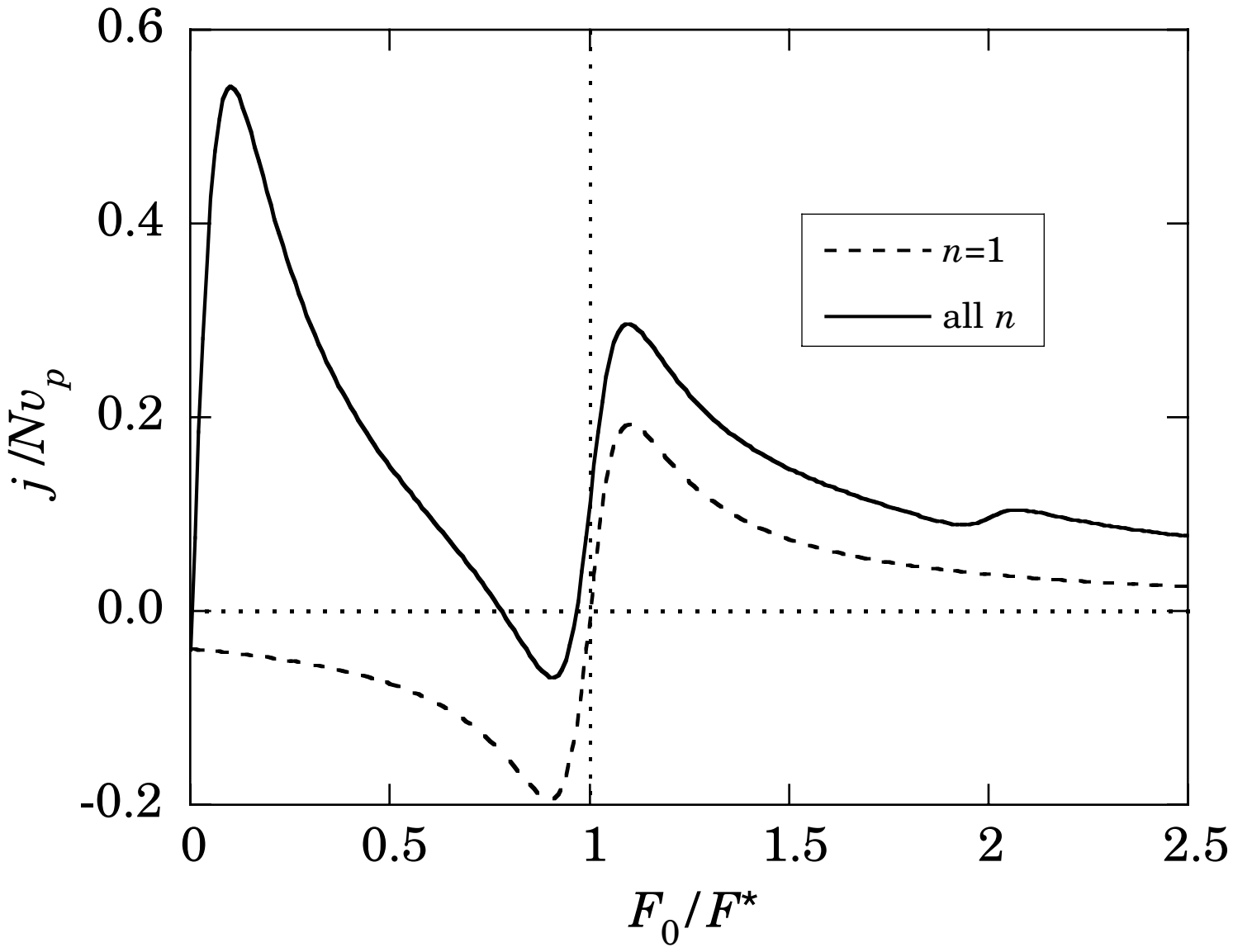

**Fig. 3.** DC current, in units of $Nv_p$, obtained by weighted summation over the photon replicas of $v(F)$, assuming the default parameters $F^*/F_c = 10$ and $F_\omega/F^* = 1$. The broken line is the contribution from the $n = 1$ term alone.

To appreciate the choice $z = 1$ as a default parameters, recall that LSA-style operation would require that the minimum reached by the overall bias field during each oscillation cycle, $F_0 - F_\omega$, falls below the critical field $F_c$ of the static $v(F)$ characteristic. Given a dc bias field in the range (30), this implies $F_\omega \approx F^*$, or $z \approx 1$. However, we shall find shortly that the useful parameter range is much wider than what this LSA analogy suggests.

It is evident from Fig. 3 that, although the $n = 0$ term in the sum makes a substantial contribution to the dc *current* in the bias range of interest, the small-signal *conductivity*, which is given by the *slope* of the curve, is dominated by the steep slope of the $n = 1$ term.

The point is elaborated on in Fig. 4, which displays the conductivity. As is indicated by the broken line, for our default parameters, the $n = 1$ term alone completely dominates the conductivity in the bias range of interest. At least for these parameters, we may therefore approximate the dc conductivity by the contribution from this term alone, which leads to

$$\sigma(0) \equiv \left(\frac{\partial j_0}{\partial F_0}\right)_{F^*} \approx \frac{2Nv_p}{F_c}\cdot\left[J_1(z)\right]^2\cdot\frac{1-\gamma^2}{\left(1+\gamma^2\right)^2}, \tag{31}$$

where

$$\gamma \equiv \left(F_0 - F^*\right)/F_c\,. \tag{32}$$

We evidently have a positive dc conductivity over much of the bias field range $F^* - F_c < F_0 < F^* + F_c$.

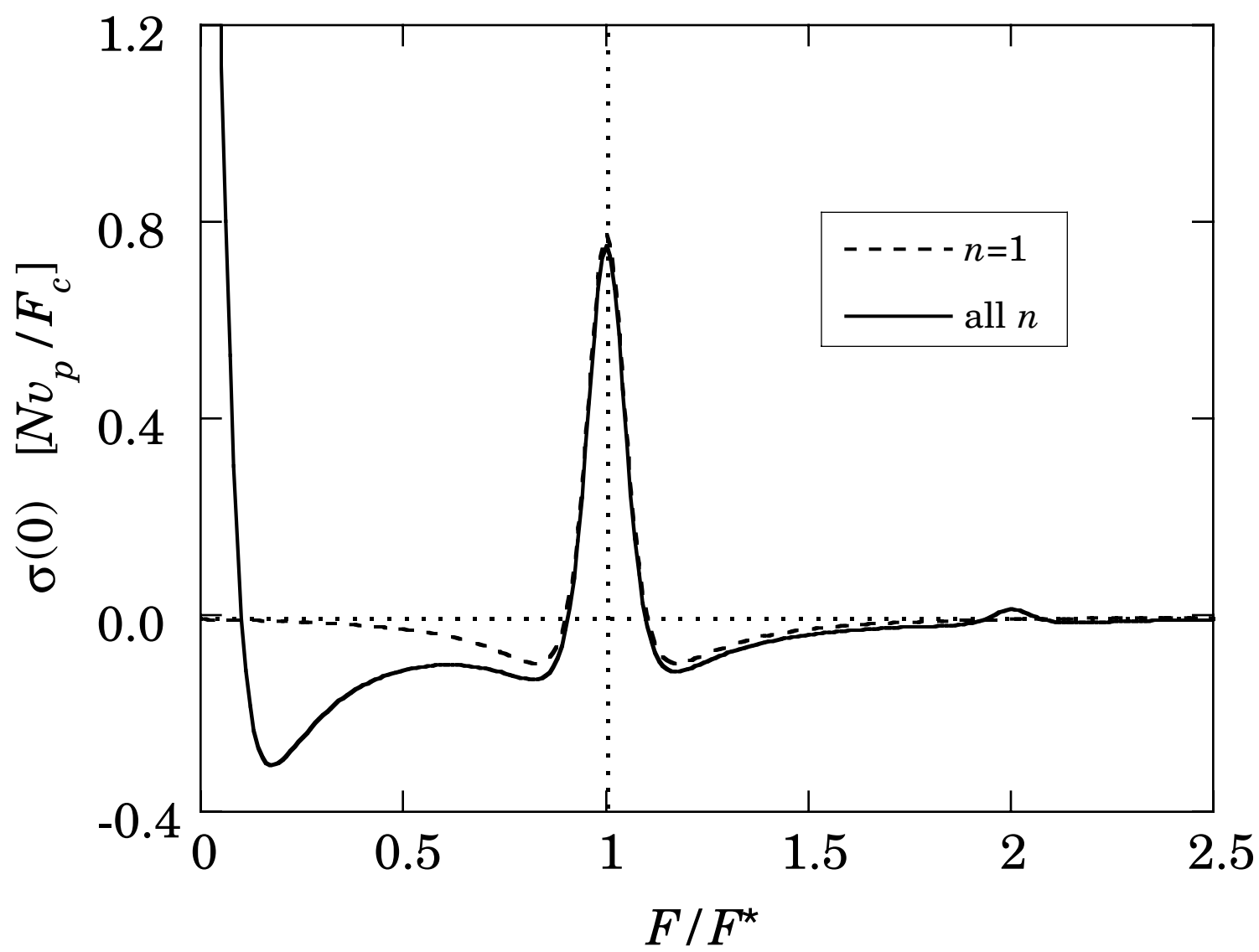

**Fig. 4.** Normalized differential dc conductivity (in units of $Nv_p/F_c$) for the same default parameters as in Fig. 3. The broken line represents the contribution from the $n = 1$ term alone; it evidently dominates the conductivity in the bias range of interest.

To estimate the effect of the neglected $n \neq 1$ terms in (27), we add the slopes of the three terms with $n = 0$, 1, and 2, all taken at the point $F_0 = F^*$, neglecting all terms further away from $n = 1$, which make a much smaller contribution. One finds that the net slope at $F_0 = F^*$ is positive when

$$\frac{\left[J_1(z)\right]^2}{\left[J_0(z)\right]^2 + \left[J_2(z)\right]^2} > \frac{\left(F^*/F_c\right)^2 - 1}{\left[\left(F^*/F_c\right)^2 + 1\right]^2}. \tag{33}$$

Consider first an $F^*$ value as low as $F^* = 2F_c$, below the expected range of interest for a practical Bloch oscillator. The right-hand side of (33) then has the value 0.120, and the inequality is satisfied for $z > 0.656$, implying a minimum ac force amplitude $F_\omega = 0.656\,F^* = 1.31\,F_c$. During each oscillation cycle, the overall drive force dips to $F_{\min} = F_0 - F_\omega$. Because of (30),

$$F^* - F_\omega < F_{\min} < F^* - F_\omega + F_c, \tag{34}$$

which for our example reduces to $0.688\,F_c < F_{\min} < 1.688\,F_c$. At least at the lower limit of this interval, $F_{\min}$ dips below the critical force of the static $v(F)$ characteristic, as one might expect from the analogy with the Gunn-LSA mode. However, with increasing $F^*$, this similarity quickly breaks down: Already for $F^* = 3F_c$, one finds $F_\omega = 0.55\,F^* = 1.64\,F_c$, and $F_{\min} > 1.36\,F_c$, clearly outside the Gunn-LSA range. In the limit $F^* >> F_c$, the minimum $F_\omega$ eventually saturates around $2F_c$, implying $F_{\min} \approx F_0 - 2F_c > F^* - 2F_c$, far above $F_c$.

This argument should not be misconstrued as denying the usefulness of an ac drive amplitude sufficiently large to cause the field to dip below $F_c$. To the contrary: Eq. (33) and the examples flowing from it only pertain to the *minimum* ac drive amplitude that is *necessary* to cause a positive dc conductivity. Larger drive amplitudes will in fact be beneficial, possibly all the way up to the maximum of $J_1(z)$, at $z \approx 1.84$, that is, $F_\omega = 1.84F^*$. For such large drive amplitudes, $F_{\min}$ would not only dip below $F_c$, as in the Gunn-LSA mode, it would dip below zero, at which point the LSA mode would no longer provide gain. Yet we will see shortly that the large-signal ac conductivity at $\omega$ remains negative over this expanded range. Evidently, the LSA analogy breaks down at *both* high and low ends of the ac field range.

The differences between the Gunn-LSA arguments and the analysis presented here are, of course, the results of a very different physics underlying the origin of the negative differential conductivity. As was elaborated in detail in [3], the NDC in the BO is the result of electron bunching in $k$-space, as reflected in our traveling-wave ansatz (8) for the distribution function. With increasing ac drive amplitude, the current associated with the traveling waves becomes increasingly non-sinusoidal, containing an increasing dc component. For a *given* ac drive amplitude, the contribution of the $n = 1$ term to the ac-drive-induced dc current vanishes when the dc bias force $F_0$ equals the characteristic field $F^*$, but it becomes positive when $F_0$ exceeds $F^*$, implying a positive differential dc conductivity.

## 4. AC currents and conductivities

### *4.1. Fundamental frequency*

The current contributions at the drive frequency $\omega$ arise from the terms with $m - n = \pm 1$ in the sum in (21), especially the cosine term in that sum, which corresponds to a real (positive or negative) conductivity; the sine-term is 90° out of phase; it represents a reactive current. Drawing again on the formulation in terms of the $v(F)$-characteristic, we write the cosine current as

$$j_{\omega}^{(c)}(t) = \left[ \frac{N}{F_{\omega}} \cdot \sum_{n=-\infty}^{+\infty} -\mathrm{J}_n \cdot \left( \mathrm{J}_{n-1} + \mathrm{J}_{n+1} \right) \cdot v\left( F_0 - nF^* \right) \right] \cdot F_{\omega} \cos(\omega t). \qquad (35)$$

The factor $F_{\omega} \cdot \cos(\omega t)$ at the end is the ac part of the drive force. Hence, the preceding square bracket represents, by definition, the (large-signal) ac conductivity at the drive frequency. Using the Bessel function identity

$$\mathrm{J}_{n-1} + \mathrm{J}_{n+1} = \frac{2n}{z} \cdot \mathrm{J}_n = \frac{2nF^*}{F_{\omega}} \cdot \mathrm{J}_n, \qquad (36)$$

we can simplify the bracket and write

$$\sigma_{\omega}^{(c)} = -\frac{2N}{F^*} \cdot \left( \frac{F^*}{F_{\omega}} \right)^2 \sum_{n=-\infty}^{+\infty} n \cdot \left[ \mathrm{J}_n(z) \right]^2 \cdot v\left( F_0 - nF^* \right). \qquad (37)$$

Fig. 5 shows the result, along with the contribution of the $\mathrm{J}_1$-term alone.

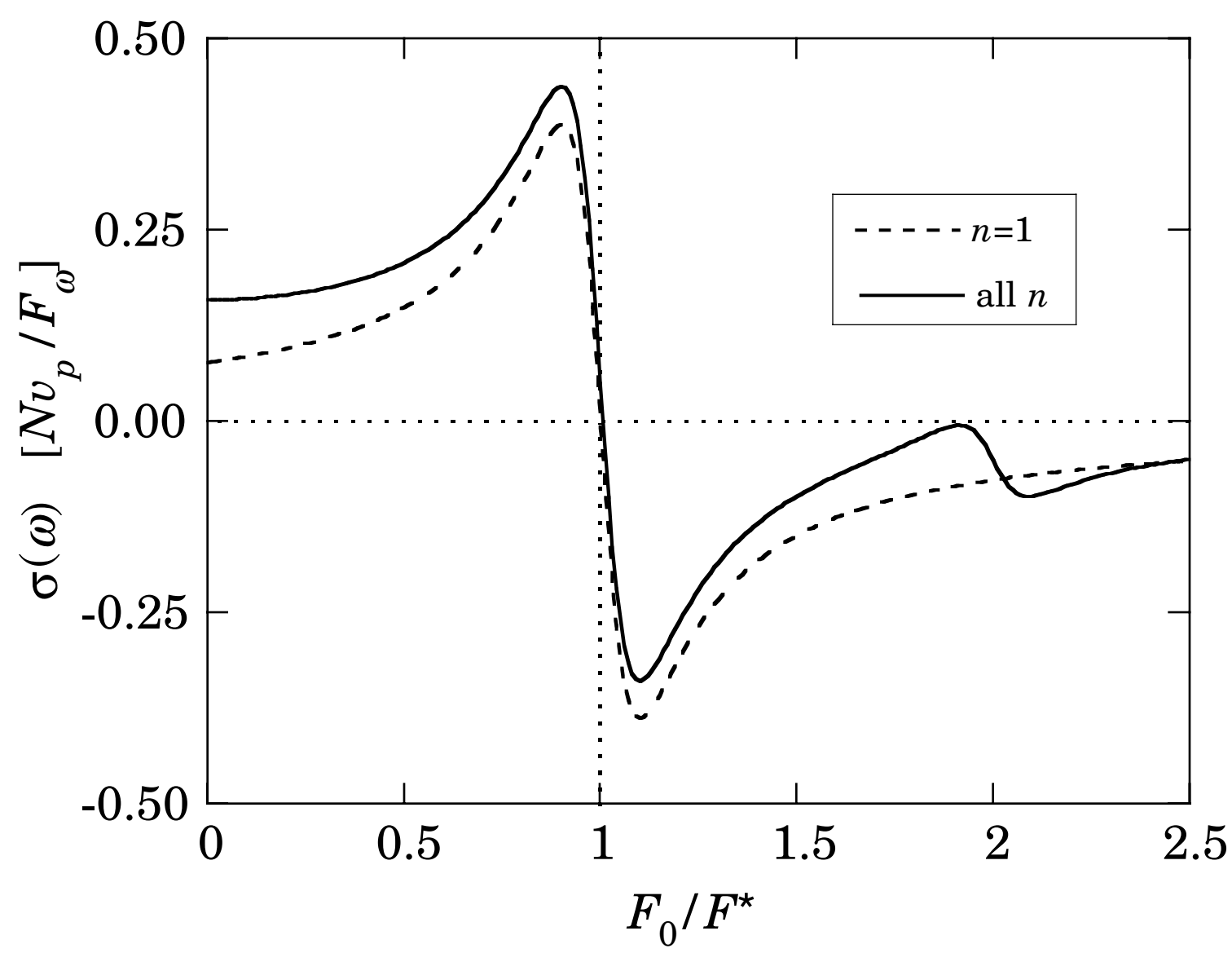

**Fig. 5.** Normalized ac conduction current at the drive frequency $\omega$, from (35), assuming the parameters of our reference sample. As in Fig. 4, the broken line represents the contribution from the $n$ = 1 term alone; it evidently dominates the conductivity in the bias range of interest. The deviations are due to “spillover” contributions, not only from the $n$ = 2 term, but also from the $n$ = –1 term.

The sum is similar to that in (27), except that now every term has the additional weighing factor $n$, which causes the previously so important $J_0$-term to drop out. For our default parameters, the latter again dominates in the bias range of interest, although not as dramatically as for the dc conductivity.

If we restrict ourselves to the $n = 1$ term, the result may be written

$$\sigma_\omega^{(c)} = -\frac{N}{2F^*}\cdot\eta(z)\cdot v\left(F_0 - F^*\right) = -\frac{Nv_p}{F^*}\cdot\eta(z)\cdot\frac{\gamma}{1+\gamma^2}, \tag{38}$$

where $\gamma$ is the same as in (32), and the dimensionless factor

$$\eta(z) = \left[\frac{2\,\mathrm{J}_1(z)}{z}\right]^2 = \left[\frac{2\,\mathrm{J}_1(F_\omega/F^*)}{F_\omega/F^*}\right]^2 \tag{39}$$

gives the dependence of the conductivity on the amplitude of the ac drive field, normalized to the value at $z = 0$, where $\eta = 1$ (Fig. 6).

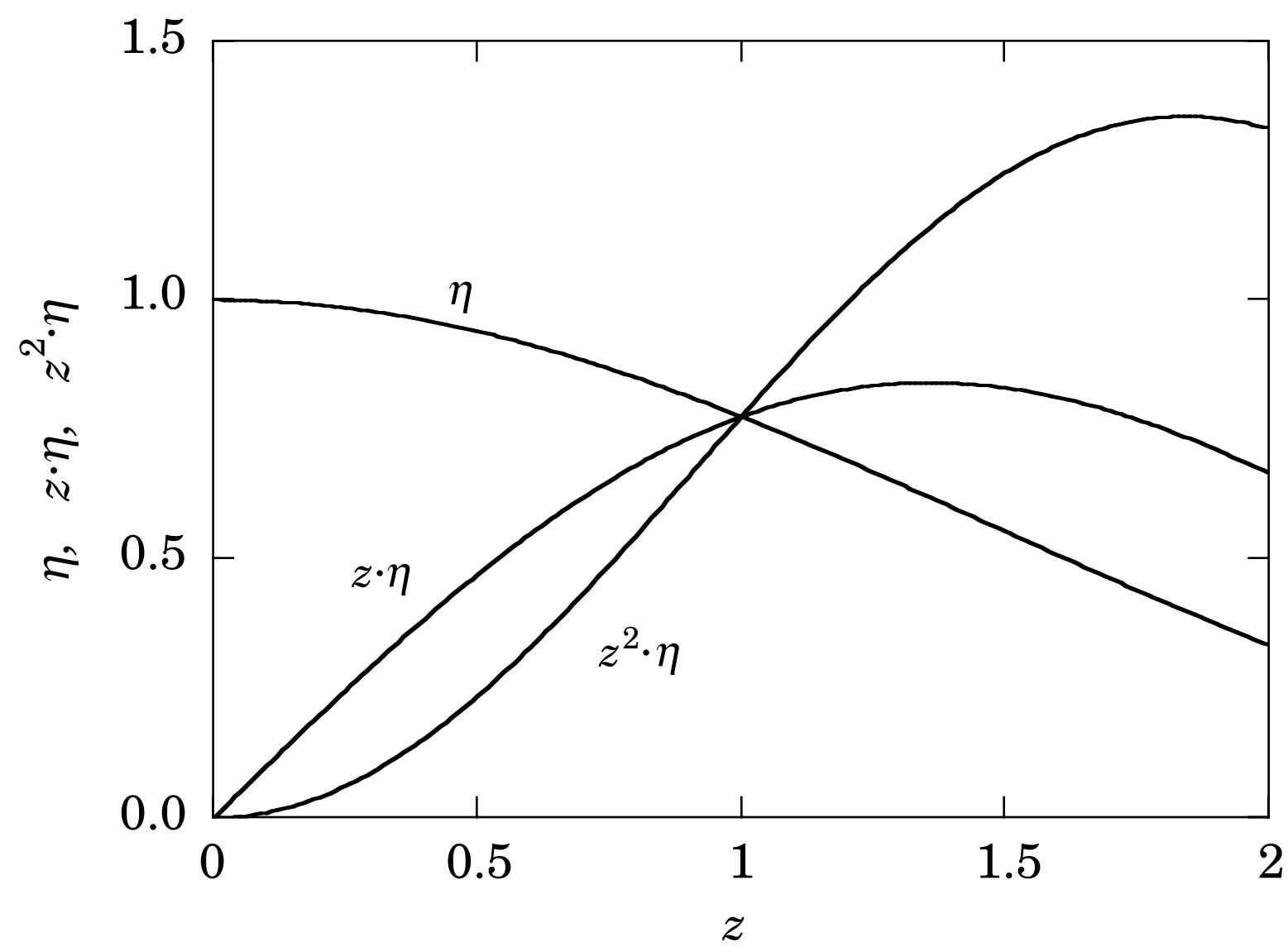

**Fig. 6.** Dependence of the large-signal ac conductivity, the ac current, and the delivered ac power, on the amplitude of the ac drive force, in units of $F^*$. The three curves plotted are $\eta(z)$, $z\cdot\eta$, and $z^2\cdot\eta$. At $z = 1$, all three functions have the value 0.77.

Because of the minus-sign in (38), the ac conductivity is negative in the interval (30) inside which the dc conductivity is positive. Although the magnitude of the negative conductivity decreases slowly with increasing drive, the ac current itself is proportional to $z\cdot\eta$; it increases with increasing drive amplitude, up to approximately $z$ = 1.35. The total ac power delivered to the resonance circuit peaks at even larger drive amplitudes; it scales with $z^2\cdot\eta$, which peaks at $z$ = 1.84. Those dependences are also shown in Fig. 6. However, for such high drive amplitudes, the single-term approximation (38) to the full sum in (37) is no longer a good approximation. We are not pursuing that matter further.

Returning to Fig. 5, note that, for our default parameters, there is a second region of negative ac conductivity just above $2F^*$. It is too weak to be of practical interest for the operation of actual BOs, especially in the face of the higher dissipation accompanying the doubled bias field.

The ratio of the small-signal-dc conductivity to the large-signal-ac conductivity is obviously of interest. From (31) and (38) we obtain

$$\left|\frac{\sigma(0)}{\sigma_\omega^{(c)}}\right| = \frac{F_\omega^2}{F^* F_c}\cdot\frac{1-\gamma^2}{2\gamma\cdot\left(1+\gamma^2\right)}. \tag{40}$$

Depending on where the dc bias field is chosen in the interval (30), this ratio can range all the way from zero (for $F_0 = F^* + F_c$) to infinity (for $F_0 = F^*$). Picking an ad-hoc value in the center of the interval, $F_0 = F^* + 0.5F_c$ ($\gamma$ = 1/2), and using our default parameters for the fields, we obtain a ratio of 6, implying a strong quenching of any low-frequency space charge instabilities.

### *4.2 Reactive currents*

The reactive current consists of two parts, an electronic contribution, and the dielectric displacement current.

The electronic part is readily obtained by a procedure analogous to that for the conductive current, but now taking the sine-part of (21), for $m - n = \pm 1$, and making everywhere the substitution

$$v(F_0 - nF^*) \rightarrow \frac{F_c}{F_0 - nF^*}\cdot v(F_0 - nF^*) = \frac{2v_p}{1+\left[\left(F_0 - nF^*\right)/F_c\right]^2}. \tag{41}$$

As in the case of the $\cos\omega t$ current, we restrict ourselves to the $n$ = 1 term, leading to

$$j_\omega^{(s)}(t) = -\frac{Nv_p}{F^*}\cdot\frac{\eta(z)}{1+\gamma^2}\cdot F_\omega \sin(\omega t). \tag{42}$$

Comparison with (38) shows that the amplitude of the reactive electron current is $1/\gamma$-times that of the conductive current. With $F_0 - F^*$ necessarily being less than $F_c$, the reactive current is larger than the conductive current, typically by about a factor 2.

The dielectric displacement current is (in our particle-current units)

$$j_{\omega}^{(d)}(t) = \frac{\varepsilon}{-e}\frac{d}{dt}\left[E_{\omega}\cos(\omega t)\right] = -\frac{\varepsilon\omega}{e^2}\cdot F_{\omega}\cdot\sin(\omega t), \tag{43}$$

where $\varepsilon$ is the dielectric permittivity of the semiconductor.

The ratio of the two reactive currents is

$$\frac{j_{\omega}^{(s)}}{j_{\omega}^{(d)}} = \frac{e^2 N v_p}{\varepsilon\omega F^*}\cdot\frac{\eta(z)}{1+\gamma^2} = \frac{1}{8}\cdot\frac{e^2 N a^2}{\varepsilon}\cdot\frac{\mathcal{E}_0}{(\hbar\omega)^2}\cdot\frac{\eta(z)}{1+\gamma^2}. \tag{44}$$

To obtain a numerical estimate, assume: $N = 1\times10^{16}\mathrm{cm}^{-3}$, $a$ = 20nm, $\varepsilon = 13\varepsilon_0$, $\mathcal{E}_0$ = 30meV, $\omega/2\pi$ = 3THz, $z$ = 1, and $\gamma$ = 0.5. They yield a ratio of 0.084. Evidently, at least for those values—which may be debatable—the displacement current would dominate.

*4.3. Higher harmonics*

The higher harmonics of the current are obtained by essentially the same procedure as the fundamental. We consider here only the cos($2\omega t$)-current, arising from the terms with $m - n = \pm 2$. Instead of (35) we now obtain

$$j_{2\omega}^{(c)}(t) = N\cdot\left[\sum_{n=-\infty}^{+\infty} + \mathrm{J}_n\cdot\left(\mathrm{J}_{n-2} + \mathrm{J}_{n+2}\right)\cdot v\left(F_0 - nF^*\right)\right]\cdot\cos(2\omega t), \tag{45}$$

Again the $n$ = 1 term is the largest,

$$j_{2\omega}^{(c)}(t) = -N\cdot\mathrm{J}_1\cdot\left(\mathrm{J}_1 - \mathrm{J}_3\right)\cdot v\left(F_0 - F^*\right)\cdot\cos(2\omega t). \tag{46}$$

For $z$-values of practical interest, $\mathrm{J}_3$ is significantly smaller than $\mathrm{J}_1$. If we simply neglect the $\mathrm{J}_3$-term, we find that the amplitude ratio of the second harmonic to the fundamental is $F_{\omega}/2F^*$, increasing proportional to the ac drive amplitude — as one might expect— but staying below unity.

Analogous results are obtained for the sin($2\omega t$)-current.

## 5. Discussion

We have obtained closed-form approximate solutions for the electron currents in a 1-D superlattice biased by both a dc force and a single-frequency ac force, with essentially arbitrary strengths of the two force components.

Except for the restriction to 1-D, the approximations necessary to obtain such a result were relatively mild: (*i*) We have neglected all higher harmonics in a $k$-space Fourier expansion of the electron distribution function—probably a fairly inconsequential simplification. (*ii*) We have used a relaxation time approximation for the scattering processes, with a single energy-independent time constant for both momentum and energy relaxation. As was discussed in [3], this may be a good approximation in a 1-D treatment, but almost certainly not in 3-D, calling for more numerical methods there, probably in the form of an extension of the Monte Carlo calculations by Anderson and Aas [15] to time-varying bias fields.

The central result of the present paper is probably the demonstration that, for a sufficiently large amplitude of the ac force, the differential dc conductivity turns positive over a certain range of dc bias values over which the ac conductivity is negative. Hence, I see a realistic chance of purely dynamic dc stabilization and domain suppression.

The ac amplitudes *necessary* to suppress the NDC at low frequencies were found to be significantly less than in the Gunn-LSA mode that stimulated the present investigation. More specifically, for dc bias fields large compared to the critical field—probably a requirement for any successful BO—the overall bias field need not dip below the critical field during every oscillation cycle, in striking contrast to the requirements for the Gunn-LSA mode.

There remains, however, the question of the “device turn-on”: The dc stabilization operates only once the ac amplitude has grown to a certain minimum value. During dc bias turn-on, this condition is initially not satisfied. Will the device automatically make the transition from the dc-unstable starting condition to the dc-stable steady-state oscillating condition? The relaxed ac amplitude requirement compared to the Gunn-LSA mode may make it easier for the oscillations to build up to the stabilizing value during startup. But low ac amplitudes also limit the ac power generation. These arguments suggests the need for special measures during turn-on, be they circuit design measures, the injection of a transient Bloch oscillation pulse, or the use of a suitably shaped turn-on ramp. Evidently, further research on this topic is needed.

Finally, I have not touched at all on what is possibly the most severe unsolved problem, the contact boundary problem. Throughout our treatment, we have assumed a spatially uniform field throughout the SL. But at the ends of the SL the field must somehow go over into the very low field inside the two contacts, with current continuity through the interface. While this would be a relatively inconsequential complication at the contacts to a positive-conductivity medium, the behavior of any negative-conductivity medium is known to depend strongly on the boundary conditions at the electrodes, especially at the negative electrode (commonly referred to as the cathode), which acts as the source of the electron flow. For example, it has been shown [16, 17] that a medium with negative differential mobility, when contacted by barrier-free conventional $n^+/n$ ohmic contacts, will nevertheless exhibit a dc current-voltage characteristic with a positive slope, as a result of the injection of excess electrons by the negative contact. Furthermore, in a well-designed Gunn diode with ohmic cathode contacts, the domains are not nucleated by statistical fluctuations in the semiconductor body, but by the large transition in field strength at the cathode-to-bulk interface [12, 18]. In an $n^+$-terminated SL biased by a pure dc voltage, one would expect a similar domain-dominated behavior, suppressing true Bloch oscillations. But it is less clear what would happen in the presence of a large ac amplitude. On the one hand, the accompanying dynamic dc stabilization deep inside the SL would probably suppress traveling domains downstream from the field transition region, and the (undesirable) transit-time oscillations accompanying them. On the other hand the dynamic stabilization works only over a relatively narrow dc bias field range, and much of the contact-to-SL transition region remains in a negative-conductivity regime, with unknown—but almost certainly undesirable—consequences. Evidently, future research on these boundary condition problems and optimum contact design is needed—there is clearly a strong incentive for doing so.

From a physics perspective, an interesting aspect of our treatment is that the entire electron dynamics can be described in terms of the photon replicas of the static $v(F)$ characteristic, with the dominant replica corresponding to the emission of a single photon. This behavior can be readily understood as follows. When we insert a negative-conductivity SL into a high-$Q$ electromagnetic resonator, we are creating a combination of two coupled quantum system that can—and will—exchange photons. Negative conductivity means that photons are transferred from the electrons to the resonator, essentially a form of stimulated emission. The energy for the transfer comes of course from the dc bias source. The necessary condition for NDC operation, $F_0 > F^*$, means that the energy $F_0 a$ picked up from the dc source along the distance of one SL period $a$ must exceed the emitted photon energy $\hbar\omega$ (see also Ignatov et al. [4] on this point). Once this condition is satisfied, photon emission can take place, but for each emission event, only the remaining energy $F_0 a - \hbar\omega$ is available for conversion into kinetic energy of the electron. The net result is the same *as if* only the force $F_0 - F^*$ were available for electron acceleration. Hence the photon replicas. A more rigorous treatment of this problem would require a fully quantized treatment of the combined electron-photon system, rather than our essentially semi-classical treatment in which the resonator is treated as a classical external potential.

## Acknowledgment

The research reported here grew out of numerous discussions with Prof. S. J. Allen on the problems that have to be solved to achieve a practical cw Bloch oscillator terahertz source.

## Appendix

The formulation (3) of Newton's Law in $k$-space applies to a time-dependent *effective wave number k*. But in a crystal of finite size, the allowed values of the Bloch wave number form a time-independent discrete set {K}. The quantity $k$ appearing in (3) then is not the (time-independent) wave number of a particular Bloch wave, but a (time-dependent) average defined as follows.

The state of the electron may be expressed as a linear superposition of Bloch waves, each belonging to one of the allowed $K$-values. Each Bloch wave is a symmetry eigenstate of the lattice translation operator $\hat{T}$,

$$\hat{T}\psi_K(x) \equiv \psi_K(x+a) = T_K \psi_K(x), \quad \text{where} \quad T_K = \exp(iKa). \qquad \text{(47a,b)}$$

Associated with a superposition state of the electron is then a certain expectation value of the symmetry eigenvalues, which may be written as the product of a real magnitude and a phase factor,

$$\langle T_K \rangle \equiv \langle \exp(iKa) \rangle = |\langle T_K \rangle| \cdot \exp[ik(t)a]. \qquad (48)$$

It can be shown [19, 20] that, in the presence of a spatially uniform force $F$, the magnitude $|\langle T_K \rangle|$ is time-independent, and the quantity $k(t)$ in the phase factor is the effective wave number of the state, in the sense that it obeys Newton's law in the form (3), including the case of a time-dependent force. In fact, this relation remains true even in the presence of transitions to higher bands. In this sense, (3) is an exact law.